\documentclass[11pt]{article}
\pdfoutput=1
\usepackage{jheppub}
\usepackage[usenames,dvipsnames,table]{xcolor}
\usepackage{graphicx,amsmath,amssymb,amsthm,multirow,array,bm,bbm,esint}
\usepackage[mathscr]{eucal}
\usepackage[bbgreekl]{mathbbol}
\usepackage{epsf,amsfonts}
\usepackage{slashed}
\usepackage[numbers,sort&compress]{natbib}
\usepackage{minibox}
\usepackage{pdflscape}
\usepackage{comment}
\usepackage{multirow,subfig}
\usepackage{mathtools}

\DeclareSymbolFontAlphabet{\mathbbm}{bbold}
\DeclareSymbolFontAlphabet{\mathbb}{AMSb}%

\usepackage{slashed}

\newcommand{\dr}{\text{\tiny dressed}}
\newcommand{\diff}{\text{Diff}$(S^1)$}
\newcommand{\diffs}{\text{Diff}$(S^1)/U(1)$}
\newcommand{\diffv}{\text{Diff}$(S^1)/SL(2)$}
\newcommand{\slt}{SL(2,\mathbb{R})}
\newcommand{\cF}{\mathcal{F}}
\newcommand{\ad}{\text{Ad}}

\def \Tr{\mbox{Tr\,}}

\def \sch{\mbox{Sch}\! }

\def\res{\mathop{Res}\limits}

\newcommand{\ie}{\emph{i.e.}, }
\newcommand{\eg}{\emph{e.g.}, }

\title{A Bound on Chaos from Stability}

\author[\, a,b,c]{Junggi Yoon}
\affiliation[\,a]{School of Physics, Korea Institute for Advanced Study\\
85 Hoegiro Dongdaemun-gu, Seoul 02455, Republic of Korea.}
\affiliation[\,b]{Asia Pacific Center for Theoretical Physics,\\77 Cheongam-ro, Nam-gu, Pohang-si, Gyeongsangbuk-do, 37673, Korea}
\affiliation[\,c]{Department of Physics, POSTECH\\ 77 Cheongam-ro, Nam-gu, Pohang-si, Gyeongsangbuk-do, 37673, Korea}
\emailAdd{junggiyoon@kias.re.kr}

\abstract{
We explore the quantum chaos of the coadjoint orbit action of diffeomorphism group of $S^1$. We study quantum fluctuation around a saddle point to evaluate the soft mode contribution to the out-of-time-ordered correlator. We show that the stability condition of the semi-classical analysis of the coadjoint orbit found in~\cite{Witten:1987ty} leads to the upper bound on the Lyapunov exponent which is identical to the bound on chaos proven in~\cite{Maldacena:2015waa}. The bound is saturated by the coadjoint orbit \diffv\ while the other stable orbit \diffs\ where the $SL(2,\mathbb{R})$ is broken to $U(1)$ has non-maximal Lyapunov exponent.

}

\begin{document}
\maketitle


\section{Introduction}
\label{sec: introduction}

Chaos is a ubiquitous phenomenon observed in the nature. One way to diagnose the chaos is the butterfly effect which is known as the sensitivity of the system on the initial condition. In quantum system, this can be quantified by a thermal expectation value of the square of the commutator of two operators $V(t)$ and $W(0)$. This measures the sensitivity of the operator $V(t)$ at time $t$ on the initial perturbation $W(0)$ at time $t=0$~\cite{Shenker:2013pqa,Shenker:2014cwa}. In chaotic system, this is expected to grow exponentially in time $t$. This exponential growth comes from the out-of-time-ordered correlator~(OTOC)\footnote{The usual OTOC form the square of the commutator of $V(t)$ and $W(0)$ is defined as 
\begin{equation}
\Tr(e^{-\beta H} V(t)W(0)V(t) W(0))
\end{equation}
In this paper, we take regularization scheme in \eqref{def: otoc} to evaluate the OTOC ~\cite{Maldacena:2015waa}.}
\begin{equation}
	\langle V(t)W(0)V(t) W(0)\rangle_\beta\,\equiv \, \Tr\big[e^{-{\beta H\over 4}} \,V(t)\,e^{-{\beta H\over 4}}\,W(0)\,e^{-{\beta H\over 4}}\,V(t) \,e^{-{\beta H\over 4}}\,W(0)\big] \ ,\label{def: otoc}
\end{equation}
which is one of terms in the expansion of the square of the commutator. Here, $\beta=T^{-1}$ is the inverse of the temperature. We normalize the OTOC by the leading disconnected diagrams $\langle V(t)V(t) \rangle_\beta \langle W(0)W(0) \rangle_\beta$ which is independent of time by time translational symmetry. The normalized OTOC of the chaotic system behaves as follow~\cite{Shenker:2013pqa,Shenker:2014cwa}.
\begin{equation}
	\cF(t)\,\equiv\, {\langle V(t)W(0)V(t) W(0)\rangle_\beta\over \langle V(t)V(t) \rangle_\beta \langle W(0)W(0) \rangle_\beta } \,=\, 1- \varepsilon e^{\lambda_L t} \ , \label{eq: general otoc}
\end{equation}
where $\varepsilon$ is inversely proportional to the logarithm of the entropy~$S$. For example, it is proportional to ${1\over N}$ and ${1\over N^2}$ in large $N$ vector and matrix models, respectively, and accordingly it is proportional to the Newton constant $G$ in the holographic dual gravity. This bound is shown to be saturated by CFTs dual to Einstein gravity~\cite{Shenker:2013pqa}. In early time, the OTOC is almost constant. As time passes, one can observe the exponential growth in the OTOC \footnote{This is valid up to the scrambling time $t_\ast\sim {1\over \lambda_L}\log {1\over \epsilon}$. Beyond the scrambling time, the $1/N$ perturbation will break down.}, and its growth rate $\lambda_L$ is called Lyapunov exponent.\footnote{Note that few-body classical chaotic systems do not always exhibit the exponential growth of OTOC~\cite{Hashimoto:2017oit}.}

\cite{Maldacena:2015waa} proved that the Lyapunov exponent of any QFT with basic constraints (\eg unitarity and causality) is bounded:
\begin{equation}
	\lambda_L \leqq {2\pi \over \beta}\ .\label{eq: chaos bound}
\end{equation}
This bound implies that there possibly exist maximally chaotic systems. Indeed, it was shown that the SYK model~\cite{Sachdev:1992fk,kitaevfirsttalk,KitaevTalks,Polchinski:2016xgd,Jevicki:2016bwu,Maldacena:2016hyu,Jevicki:2016ito}, the tensor model~\cite{Gurau:2010ba,Carrozza:2015adg,Witten:2016iux,Gurau:2016lzk,Klebanov:2016xxf}, the 2D dilaton gravity on nearly-AdS$_2$~\cite{Maldacena:2016upp} and string worldsheet theories~\cite{deBoer:2017xdk} are maximally chaotic. By recovering all dimensionful parameters, the maximum value of Lyapunov exponent becomes ${2\pi k_B \over \hbar \beta}$, and it diverges in the classical limit. For this reason, \eqref{eq: chaos bound} can be viewed as the bound on quantum chaos.

In this paper, we will show that the bound on quantum chaos can also be derived from the stability of semi-classical analysis of Schwarzian theory. Unlike \cite{Maldacena:2015waa} of which the proof of the bound on chaos is valid for any QFT with unitarity and causality, our discussion is restricted to the Schwarzian theory and is not as general as that of \cite{Maldacena:2015waa}. Nevertheless, our analysis provides a very simple physical argument for the bound, which would reveal the underlying physical meaning of the bound on chaos.

\section{Schwarzian Theory and Coadjoint Orbit Action}
\label{sec: schwarzian theory and coadjoint orbit action}

In SYK-like model and 2D dilaton gravity, the reparametrization symmetry along the thermal circle is broken to $SL(2,\mathbb{R})$, which leads to the pseudo-Goldstone boson described by the Schwarzian low energy effective action:
\begin{equation}
	S= \int_0^{2\pi} d\tau \;\left( -{c\over 12 } \sch\left[ \phi(\tau),\tau\right] -{c\over 24} [\phi'(\tau)]^2\right)\ .\label{eq: schwarzian action}
\end{equation}
Here, $\phi(\tau)\in$\diffv\ is a diffeomorphism of the circle modulo $\slt$.

Now, let us consider the case where the reparametrization symmetry is broken to $U(1)\subset SL(2,\mathbb{R})$. For example, such a pattern of symmetry breaking has been observed in the generalized SYK-like models in~\cite{Ferrari:2019ogc} and the 2D dilaton gravity with defects in~\cite{Anninos:2018svg,Mertens:2019tcm}. For their low energy effective action, one can include not only Schwarzian derivatives but also any possible terms which vanish under the corresponding $U(1)$ mode. Hence, the simplest effective action can be written as
\begin{align}
	S=& \int d\tau \;\left( -{c\over 12 } \sch\left[ \phi(\tau),\tau\right] + b_0 [\phi'(\tau)]^2\right)\ ,\label{eq: coadjoint orbit action}
\end{align}
where $b_0$ is a constant. This is nothing but the time-translation generator $L_0$ of a coadjoint orbit $\phi(\tau)\in \text{Diff}/H$ where $H$ is the stabilizer subgroup~\cite{Witten:1987ty,Stanford:2017thb}.

We will briefly review the coadjoint representation and its time-translation generator~\cite{Witten:1987ty,Alekseev:1988ce,Rai:1989js,Stanford:2017thb,Cotler:2018zff}. Let us consider the element $(v(\tau), a)$ of the central extension of the \diff which corresponds to Virasoro group. Here, $a$ is the central charge. The dual vector $(b(\tau),c)$ of the central extension of \diff\ can be induced via the inner product defined by
\begin{equation}
	\langle (b(\tau),c) , (v(\tau),a) \rangle \equiv \int_0^{2\pi} {d\tau \over 2\pi } b(\tau) v(\tau)+ ca\ .\label{eq: inner product}
\end{equation} 
The adjoint representation of $\phi\in$\diff\ on the vector $(v(\tau),a)$ is defined by~\cite{Witten:1987ty,Alekseev:1988ce,Rai:1989js}
\begin{align}
	\ad_{\phi^{-1}}(v(\tau),a)\,=\,\left({v(\phi(\tau))\over \phi'(\tau)} , a+ {1\over 12} \int {d\tau\over 2\pi} {v(\phi(\tau))\over \phi'(\tau)}\sch[\phi(\tau),\tau] \right)\ .\label{eq: adjoint action}
\end{align}
Via the inner product in \eqref{eq: inner product}, the adjoint representation \eqref{eq: adjoint action} induces the coadjoint representation on the dual vector $(b(\tau),c)$. In particular, we are interested in a coadjoint orbit of a constant dual vector $(b_0,c)$~\cite{Witten:1987ty,Stanford:2017thb}:
\begin{equation}
	\ad^\ast_{\phi^{-1}}[(b_0,c)] = \left(\big[\phi'(\tau)\big]^2 b_0 -{c\over 12} \sch\left[ \phi(\tau),\tau\right],c \right)\ .
\end{equation}
Depending on the constant value of $b_0$, the stabilizer of the coadjoint representation $\ad^\ast_{\phi^{-1}}$ is different. For example, for $b_0\ne -{c n^2\over 24}$ ($n\in \mathbb{N}$), the coadjoint representation $\ad^\ast_{\phi^{-1}}$ is invariant under the $U(1)$ transformation of $\phi(\tau)\;\;\rightarrow\;\; \phi(\tau)+ a$ ($a$: constant). On the other hand, the stabilizer subgroup is $SL(2,\mathbb{R})$ for $b_0= -{c \over 24}$. Hence, the coadjoint orbit is isomorphic to \diffs\ or \diffv\ for $b_0\ne -{c n^2\over 24}$ or $b_0= -{c \over 24}$, respectively.

The Hamiltonian corresponding to the Virasoro generator $L_0=i {\partial \over \partial \tau}$ can be obtained from the inner product of $\ad^\ast_{\phi^{-1}}(b(\tau)) $ and a constant vector~\cite{Witten:1987ty}. This leads to the Schwarzian action in \eqref{eq: coadjoint orbit action}.

In large $c>0$, one can perform the semi-classical analysis of the Schwarzian action in \eqref{eq: coadjoint orbit action} by studying the quantum fluctuation around the identity diffeomorphism $\phi(\tau)=\tau$. The quadratic action in the large $c$ expansion of the action reads
\begin{equation}
	S^{(2)} \,\sim \, \sum_{n}  n^2\left(n^2 + {24 \over c}b_0 \right)\epsilon_{-n}\epsilon_n\ .
\end{equation}
Therefore the fluctuation $\epsilon_n$ at the quadratic level (in particular, $\epsilon_1$) is stable if
\begin{equation}
	b_0\geqq  - {c\over 24}\ . \label{eq: stability bound}
\end{equation}
Note that this bound is saturated by \diffv. Otherwise, the stable coadjoint orbit corresponds to \diffs~\cite{Witten:1987ty}. This bound comes from the zero $n=\sqrt{-{24\over c}b_0}$ of the quadratic action which also play a crucial role in the Lyapunov exponent of OTOC.

\section{Out-of-time-ordered Correlators}
\label{sec: otoc}

Now, we will evaluate the contribution of the Schwarzian mode to OTOC. As in the SYK-like model or the dilaton gravity on nearly-AdS$_2$, we assume that the contribution of the Schwarzian mode dominates the four point OTOC. We first evaluate the contribution of the Schwarzian mode to the Euclidean four point function. Then, we will take analytic continuation to a real-time OTOC.

\subsection{Propagator of Soft Mode}
\label{sec: Propagator of Soft Mode}

In large $c$, we expand the diffeomorphism $\phi(\tau)$ around the identity $\phi(\tau)=\tau$:
\begin{equation}
	\phi(\tau)=\tau +{1\over \sqrt{c}} \sum_{|n|\geqq n_0} \epsilon_n e^{-i n\tau}\ .\label{eq: soft mode expansion}
\end{equation}
Here, we define $n_0= 1$ or $n_0=2$ for the case of \diffs\ or \diffv, respectively. Note that $|n|<n_0$ corresponds to the stabilizer subgroup of the coadjoint representation. The soft mode expansion in~\eqref{eq: soft mode expansion} gives the quadratic action of \eqref{eq: coadjoint orbit action}:
\begin{equation}
	S^{(2)}= {\pi \over 12 }   \sum_{|n| >n_0}  n^2\left(n^2 + {24 \over c}b_0 \right)\epsilon_{-n}\epsilon_n\ .
\end{equation}
From the quadratic action, one can easily read off the propagator of the soft mode $\epsilon_n$
\begin{equation}
	\langle \epsilon_{-n} \epsilon_n \rangle = {  {6\over \pi} \over n^2 \left(n^2 +{24 \over c} b_0 \right)} \qquad\mbox{for}\quad |n|>n_0\ .\label{eq: propagator}
\end{equation}
%

\subsection{Dressed Bi-local Field}
\label{sec: Dressed Bi-local Field}

In order to evaluate OTOC, one need a ``matter'' field and its four point function. In SYK model, the four point function of the fundamental fermion $\chi^i(\tau)$ ($i=1,2,\cdots, N)$ can be evaluated as a two point function of bi-local field $\Psi(\tau_1,\tau_2)\equiv {1\over N} \sum_{i=1}^N \chi^i(\tau_1)\chi^i(\tau_2)$~\cite{Jevicki:2016bwu,Maldacena:2016hyu}. Also, for the case of the nearly-AdS$_2$, \cite{Maldacena:2016upp} evaluated the OTOC by studying the two point function of bi-locals coming from a scalar field in the nearly-AdS$_2$.

One essential ingredient of the OTOC calculation is the infinitesimal transformation of the two point function of the matter field under the symmetry related to the soft mode~\cite{Maldacena:2016hyu,Yoon:2017nig,Narayan:2019ove,Jahnke:2019gxr}. For example, one need to calculate the reparametrization transformation of the two point function in SYK model or its holographic dual near-AdS$_2$. Such a transformation gives us the coupling between matter and soft mode in the soft mode channel of OTOC. In general, soft modes require the transformation of the two point function of matters under the corresponding symmetry. However, in the field theory such a transformation is not always known unlike reparametrization symmetry. For instance, the transformation of the two point function under the $\mathcal{W}_N$ symmetry, which is involved with the higher spin soft modes, is still not fully understood in the field theory. On the other hand, such a transformation can easily be evaluated holographically by a (boundary-to-boundary) Wilson line of the gauge field in $SL(2)$ BF theory for AdS$_2$~\cite{Blommaert:2018oro,Lam:2018pvp,Blommaert:2018iqz,Mertens:2019tcm,Iliesiu:2019xuh} or Chern-Simons gravity for AdS$_3$~\cite{Jahnke:2019gxr,Narayan:2019ove}. The boundary-to-boundary Wilson line in the AdS$_2$ bulk can reproduce the two point function~\cite{deBoer:2013vca,Ammon:2013hba,Castro:2018srf,Narayan:2019ove} of its holographic dual field theory, and it is gauge invariant except for the end points. In particular, the Wilson line out of a general gauge field $A$ of $SL(2)$ BF theory or Chern-Simons gravity can be understood as gravitationally dressed two point function~\cite{Jahnke:2019gxr,Narayan:2019ove}. For more general BF or Chern-Simons gravity, one can consider a small fluctuation around a fixed background gauge field corresponding to the AdS background, which leads to the expansion of the gravitationally dressed Wilson line with respect to a small fluctuation (See \cite{Jahnke:2019gxr,Narayan:2019ove} for the details). In the small fluctuation expansion, the leading term reproduces the boundary-to-boundary two point function in the fixed AdS background in the bulk theory which corresponds to the two point function in the dual field theory. In addition, the first correction term exactly gives the infinitesimal transformation of the two point function of the dual field theory under symmetry generated by the boundary soft modes. Note that the small fluctuation corresponds to the small fluctuation corresponds to soft mode living on the boundary of AdS$_2$. For the case of $SL(2)$ BF gravity, it is governed by Schwarzian boundary action for the boundary graviton~\cite{Maldacena:2016upp,Cvetic:2016eiv,Mandal:2017thl,Grumiller:2017qao,Castro:2018ffi,Cotler:2018zff,Jahnke:2019gxr,Narayan:2019ove}. This provides the physical intuition why we need such a transformation for OTOC calculations.

In order for the contribution of the soft mode $\epsilon_n$ to the four point function, we consider the two point function of the (boundary-to-boundary) Wilson lines of a gauge field $A$ which can be understood as the bi-local field dressed by the soft mode~\cite{Maldacena:2016upp,Jahnke:2019gxr,Narayan:2019ove}.
\begin{equation}
	\langle \Phi^\dr (\tau_1,\tau_2)\Phi^\dr (\tau_3,\tau_4) \rangle\ .\label{eq: four point function}
\end{equation}
\begin{figure}[t!]
\centering
\subfloat[\small Thermal Circle\label{fig: thermal circle}]{%
  \includegraphics[width=.32\linewidth]{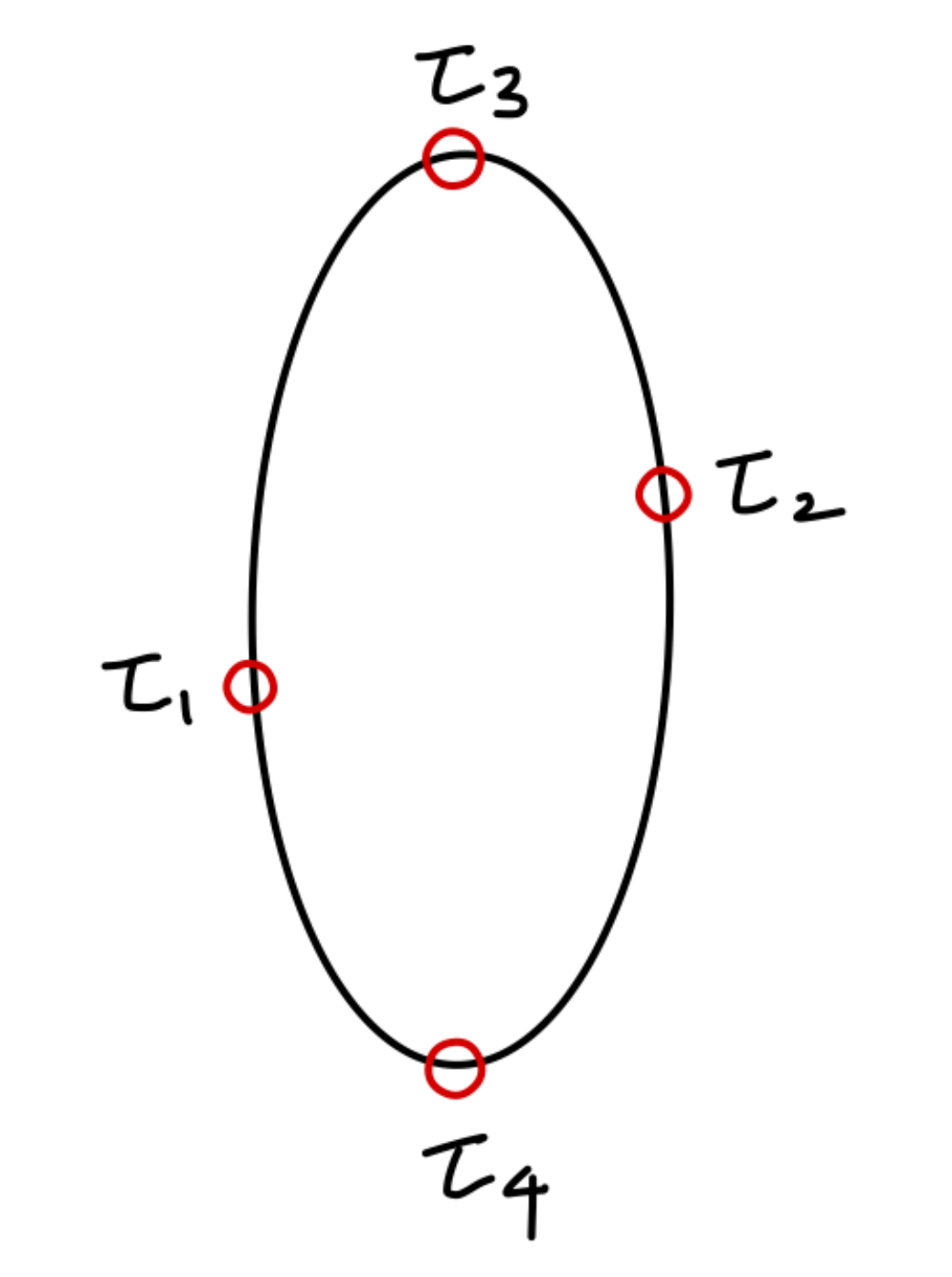}%
}\hfill \begin{minipage}[c]{.1\textwidth}
\centering
\vspace{-3.2cm}
$ \xRightarrow[\text{continuation}]{\text{analytic}}$
\end{minipage}
 \hfill
\subfloat[\small Real-time OTO\label{fig: oto}]{%
  \includegraphics[width=.46\linewidth]{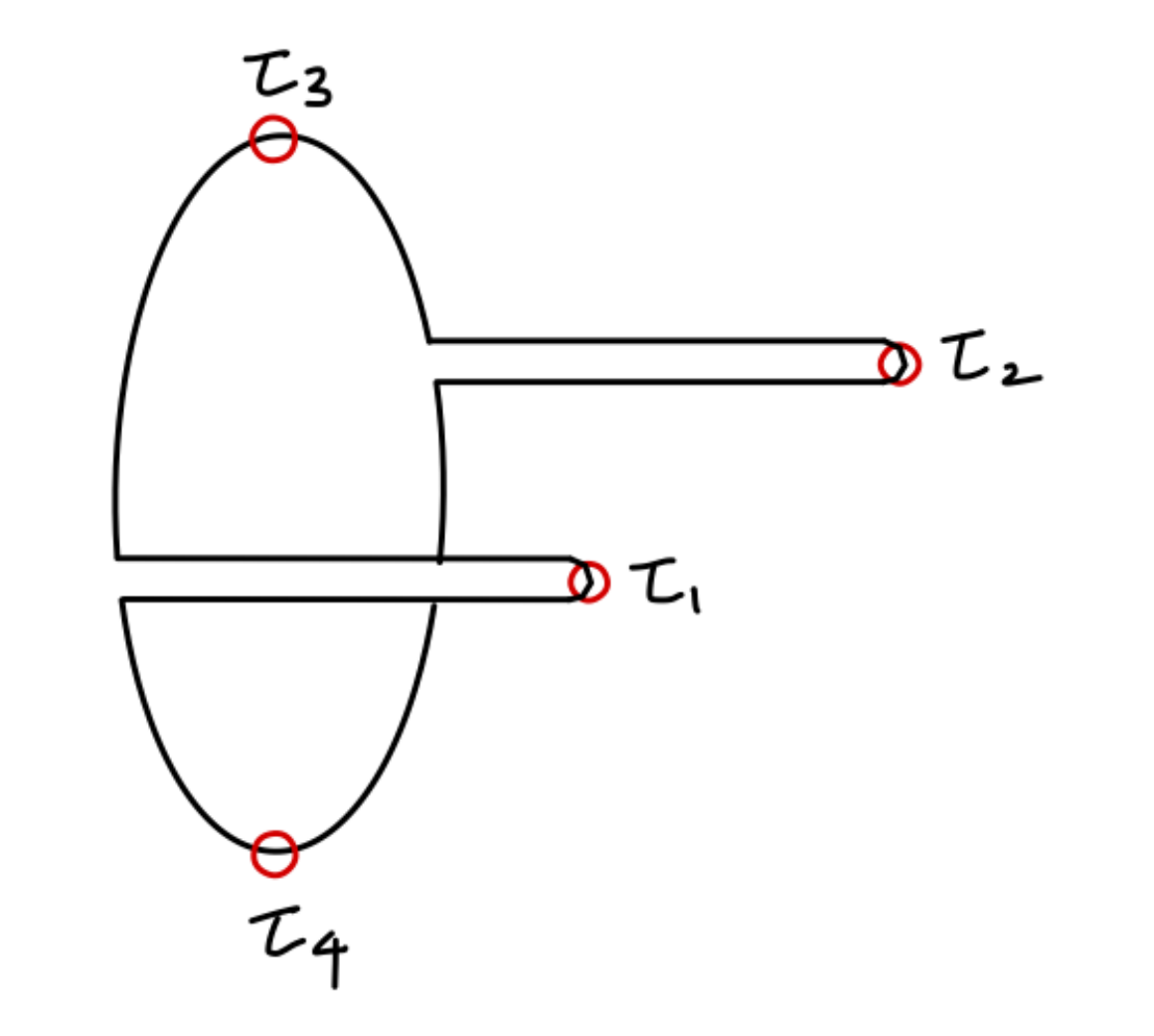}%
}
\vspace{-2mm}
\caption{We evaluate the Euclidean four point function for the configuration in \eqref{eq: otoc configuration}. Then, we analytic-continue it to real-time OTOC by \eqref{eq: analytic continuation}.}
\label{fig: configuration}
\end{figure}
%
%
For the real-time OTOC, we first evaluate the Euclidean correlator with the configuration along the thermal circle
\begin{align}
	(\tau_1,\tau_2,\tau_3,\tau_4)=(\chi-{\pi \over 2},\chi+{\pi\over 2},0, \pi  )\ ,\label{eq: otoc configuration}
\end{align}
where $\chi\in(-\pi/2,\pi/2)$ (See Figure~\ref{fig: thermal circle})~\cite{Maldacena:2016hyu,Maldacena:2016upp}. Then, later we will perform the analytic continuation of the Euclidean time $\chi$ to a real time $t$.

In large $c$, one can expand the dressed bi-local field with respect to the soft mode $\epsilon_n$: 
\begin{align}
	{\Phi^\dr(\tau_1,\tau_2)\over \Phi_{cl}(\tau_1,\tau_2) }\,=\,  1 + \sum_{|n|\geqq n_0} {\epsilon_n\over \sqrt{c} }{\delta_{\epsilon_n}\Phi^\dr(\tau_1,\tau_2)\over \Phi_{cl}(\tau_1,\tau_2) } +\mathcal{O}(c^{-1})\ ,\label{eq: soft mode expansion bilocal}
\end{align}
where $\Phi_{cl}(\tau_1,\tau_2)$ is the leading term of the soft mode expansion of the dressed bi-local, and it corresponds to the two point function in the constant background. \eg $\Phi_{cl}(\tau_1,\tau_2)\!\sim \!{1\over |\sin{\pi \tau_{12}\over \beta}|^{2h}}$. Recall that both \diffv\ and \diffs\ has $U(1)$ time translational symmetry. From the (center of) time translational symmetry, one can write the first order soft mode expansion in \eqref{eq: soft mode expansion bilocal} as follow.
\begin{equation}
	{\delta_{\epsilon_n}\Phi^\dr(\tau_1,\tau_2)\over \Phi_{cl}(\tau_1,\tau_2) } = e^{-i n \chi} f_n(\sigma) \ , \label{eq: soft mode eigenfunction}
\end{equation}
where we define the center of time and the relative time by
\begin{equation}
	\chi\equiv{1\over 2}(\tau_1+\tau_2)\quad,\quad \sigma\equiv {1\over 2} (\tau_1-\tau_2)\ ,
\end{equation}
respectively. Also, note that $\delta_{\epsilon_n}\Phi^\dr(\tau_1,\tau_2)$ can be obtained from  the infinitesimal conformal transformation of two point function because the soft mode generates the conformal transformation.

The soft mode expansion of the dressed bi-locals gives the  $1/c$ contribution of the Schwarzian soft mode to the Euclidean four point function, or equivalently two point function of the dressed bi-locals,  in \eqref{eq: four point function}. I.e.,
\begin{align}
	&\cF(\chi)\equiv {\langle \Phi^\dr (\chi-\pi/2,\chi+\pi/2)\Phi^\dr (0,\pi) \rangle\over \Phi_{cl}(\chi-\pi/2,\chi+\pi/2)\Phi_{cl}(0,\pi)}\cr
	\,=\,&1+ {1\over c}\sum_{|n|\geqq n_0} \langle\epsilon_{-n} \epsilon_n \rangle e^{in\chi } e^{-{n\pi i \over 2}}f_{-n}(-{\pi\over 2} )f_{n}(-{\pi\over2} ) \ .\label{eq: soft mode expansion of four poin function}
\end{align}
Here, we used the configuration in \eqref{eq: otoc configuration}. We will calculate the soft mode contribution for \diffv\ case and \diffs\ case, separately.

\noindent \emph{3.2 \diffv\ case.} For the case of \diffv, the soft mode eigenfunction $f(\sigma)$ is found to be~\cite{Maldacena:2016hyu}
\begin{equation}
	f_n(\sigma)\,=\,- 2 i  h  \left[n\cos n\sigma -{\sin n\sigma \over \tan \sigma } \right] \ ,
\end{equation}
where $h$ is the conformal dimension of the matter field. Note that the $SL(2,\mathbb{R})$ invariance of the bi-local fields (\eg boundary-to-boundary Wilson line or conformal two point function) leads to
\begin{equation}
	f_n(\sigma)\,=\,0\hspace{8mm}\mbox{for}\quad n=0,\pm1\ .
\end{equation}
\begin{figure}[t!]
\centering
\subfloat[\label{fig: contour integral1}]{%
  \includegraphics[width=.42\linewidth]{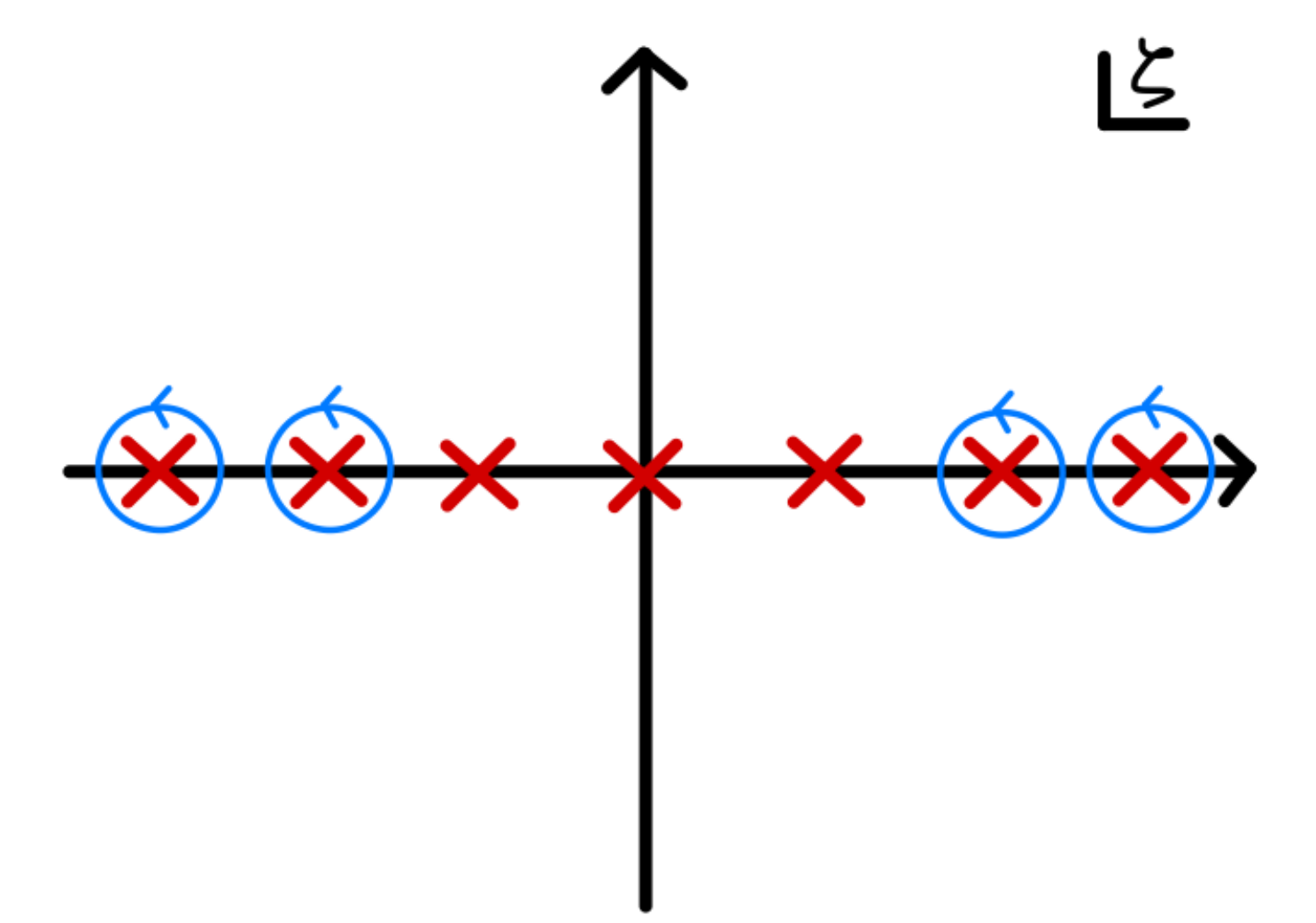}%
}\hfill \begin{minipage}[c]{.06\textwidth}
\centering
\vspace{-2.2cm}
$ \xRightarrow[\text{contour}]{\text{deform}}$
\end{minipage}
 \hfill
\subfloat[\label{fig: contour integral2}]{%
  \includegraphics[width=.42\linewidth]{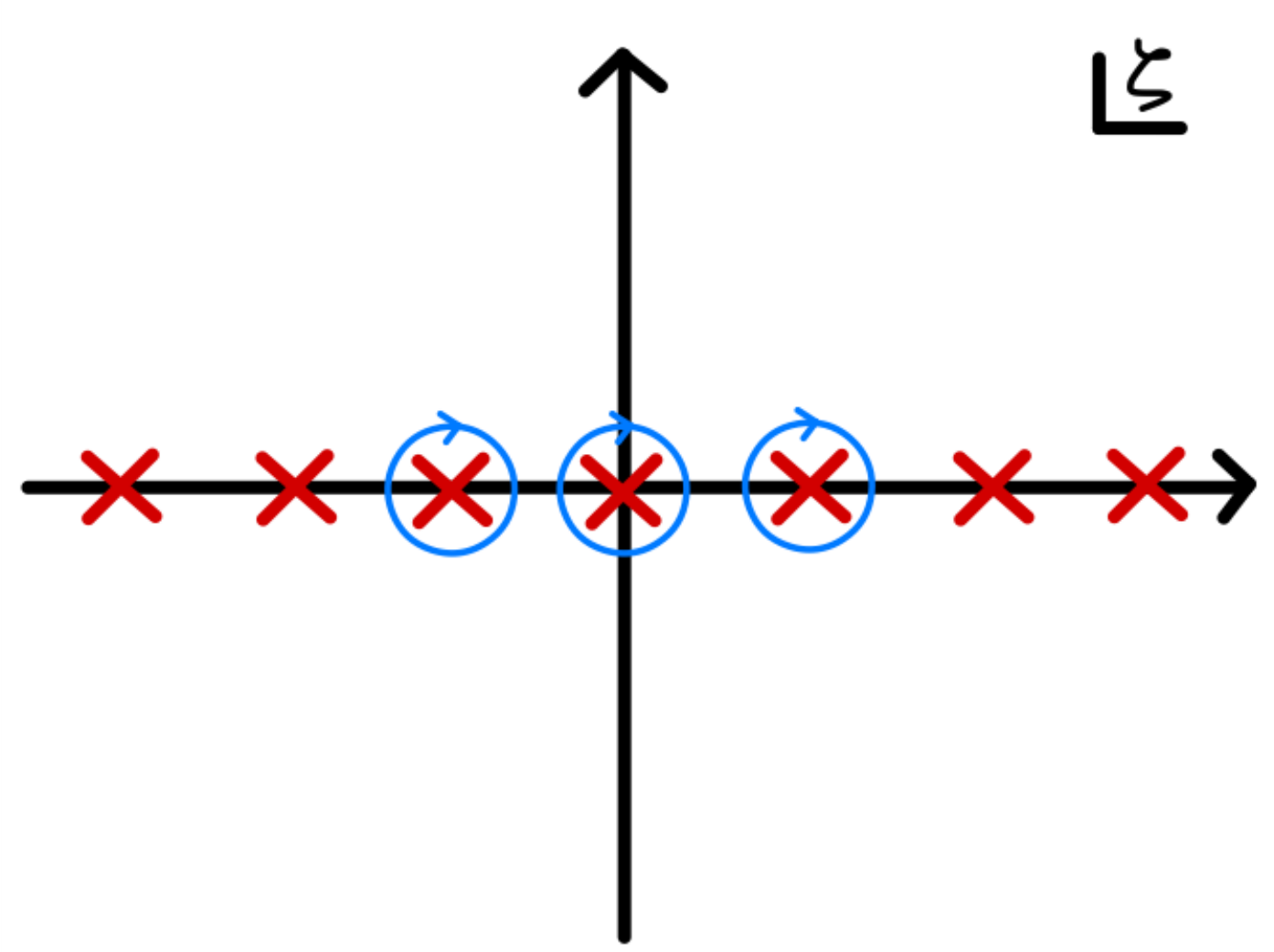}%
}
\vspace{-2mm}
\caption{The soft mode contribution can be written as a contour integral along a collection of small counterclockwise circles centered at $\zeta\in \mathbb{Z}/\{-1,0,1\}$. By deforming the contour, it can be expressed as the residue at $\zeta=-1,0,1$.}
\label{fig: contour1}
\end{figure}
Together with the propagator of soft mode in \eqref{eq: propagator}, the soft mode expansion in \eqref{eq: soft mode expansion of four poin function} can be written as a contour integral as follow~\cite{Narayan:2017qtw,Yoon:2017nig,Narayan:2017hvh,Jahnke:2019gxr,Narayan:2019ove}.
%
%
%
\begin{align}
	&\cF(\chi)=1+{24h^2\over \pi c}  {1\over 2\pi i}\oint_{\mathcal{C}} d\zeta\; { {\pi \over 2}\over \sin {\pi \zeta\over 2} } {e^{i\zeta\chi }\over \zeta^2-1}  \ ,
\end{align}
where the contour $\mathcal{C}$ is a collection of small counterclockwise circles around $\zeta=n$ ($n\in \mathbb{Z}/\{-1,0,1\}$). See Figure~\ref{fig: contour integral1}. Then, by deforming the contour, one can change it into a sum of residues at $\zeta=-1,0,1$ (See Figure~\ref{fig: contour integral2})~\cite{Maldacena:2016hyu,Maldacena:2016upp,Sarosi:2017ykf,Jahnke:2019gxr,Narayan:2019ove}:
\begin{align}
	\cF(\chi)\,=\,&1-{24 h^2\over \pi c}  \sum_{n=0,\pm 1 }\res_{\zeta=n} \; { {\pi \over 2}\over \sin {\pi \zeta\over 2} } {e^{i\zeta\chi }\over \zeta^2-1} \cr
	\,=\,&1+{24 h^2\over \pi  c } \left(1- {\pi \over 2}\cos \chi\right)\ .
\end{align}
For the real-time OTOC, we perform the analytic continuation of the Euclidean time $\chi$ to real time $t$ (See Figure~\ref{fig: oto}). \ie
\begin{equation}
	\chi\quad\longrightarrow \quad -{2\pi i \over \beta}t\ . \label{eq: analytic continuation}
\end{equation}
The analytic continuation gives the exponential growth with the maximal Lyapunov exponent $\lambda_L= {2\pi \over \beta}$~\cite{Maldacena:2016hyu,Maldacena:2016upp,Narayan:2017qtw,Yoon:2017nig,Narayan:2017hvh,Jahnke:2019gxr}.
\begin{equation}
	\cF(t)\,=\, 1- {6 h^2\over  c}e^{{2\pi \over \beta }t}+\cdots\ ,
\end{equation}
where we omit the term does not grow exponentially in time at $\mathcal{O}(c^{-1})$ order.

\subsection{\diffs\ case}
\label{sec: diff case}

We repeat a similar calculation for the \diffs\ case where $SL(2,\mathbb{R})$ is broken to $U(1)$. In this case, we also assume a matter field coupled to \diffs. And, its boundary-to-boundary two point function has $U(1)$ invariance. The infinitesimal transformation of the boundary-to-boundary correlation function will give the soft mode eigenfunction $f_n(\sigma)$. The form of the function $f_n(\sigma)$ in \eqref{eq: soft mode eigenfunction} might depend on the details of model.\footnote{The space of $U(1)$ invariant function is larger than that of $SL(2)$, one would not be able to use a general form of observables in the $SL(2)$ case. I would like to thank the referee to point out this.} Nevertheless, the evaluation of the Lyapunov exponent can be independent of the form of the function $f_n(\sigma)$ because the calculation mainly uses the center of time translational symmetry and the structure of poles. In general, the contribution of the soft mode in \eqref{eq: soft mode expansion of four poin function} now includes $n=\pm 1$ terms because they do not belong to the stabilizer subgroup for the case of \diffs. Moreover, it might be possible that $f_n(-{\pi\over 2})=0$ for odd $n$ as in \diffv. However, one can repeat the same calculation and it does not change the conclusion.

\begin{figure}[t!]
\centering
\subfloat[\label{fig: contour integral3}]{%
  \includegraphics[width=.42\linewidth]{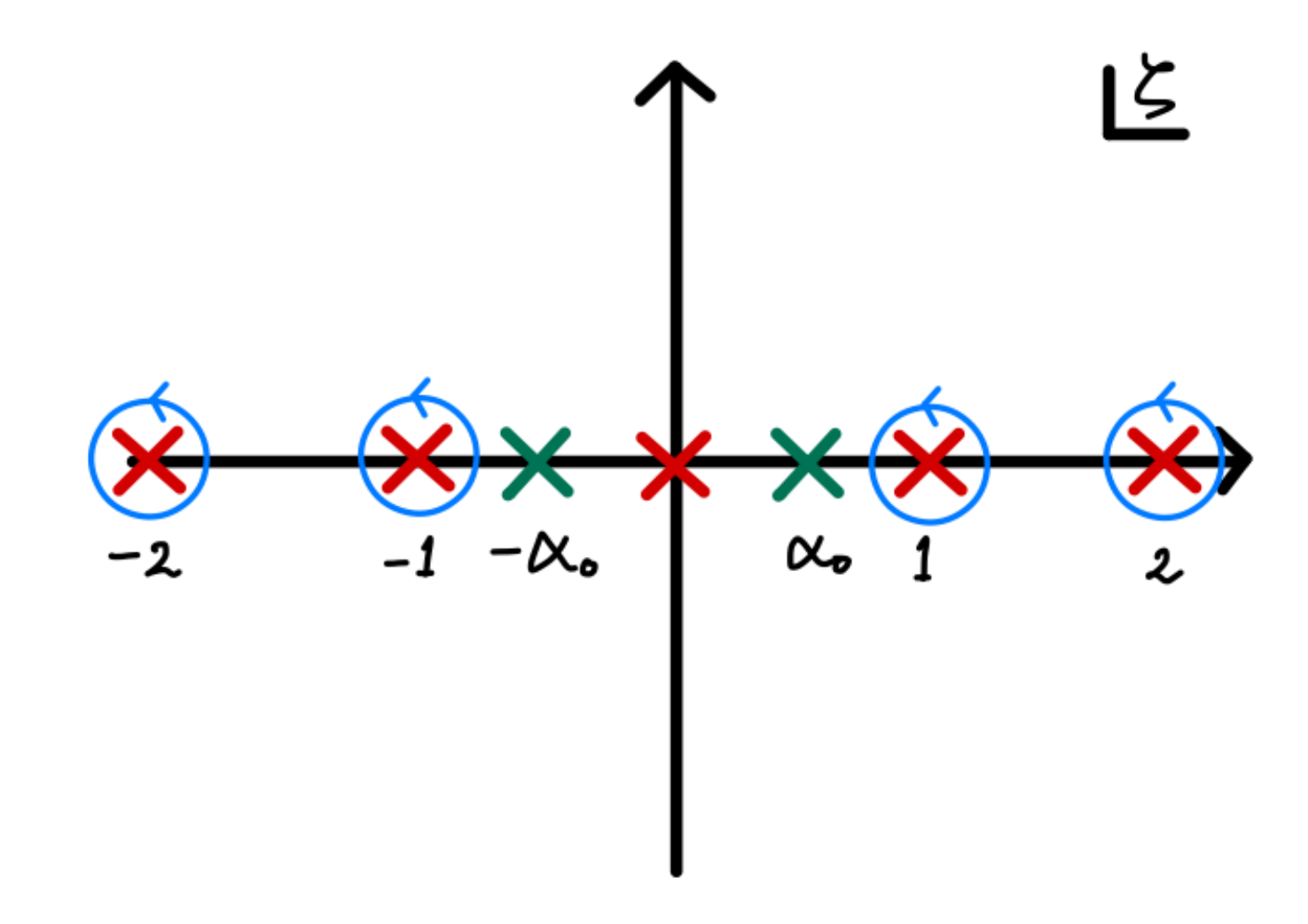}%
}\hfill \begin{minipage}[c]{.06\textwidth}
\centering
\vspace{-2.2cm}
$ \xRightarrow[\text{contour}]{\text{deform}}$
\end{minipage}
 \hfill
\subfloat[\label{fig: contour integral4}]{%
  \includegraphics[width=.42\linewidth]{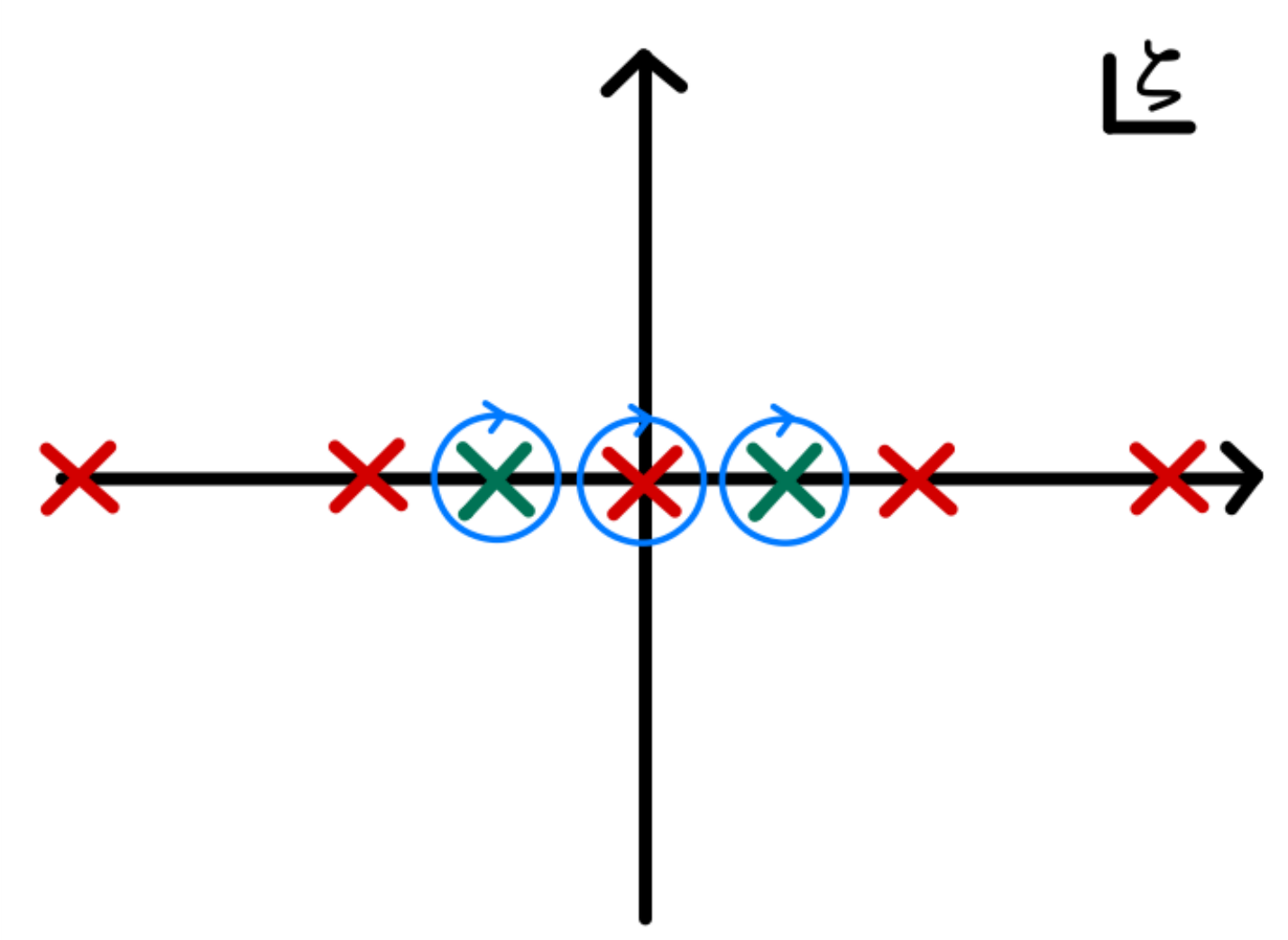}%
}
\vspace{-2mm}
\caption{The soft mode contribution can be written as a contour integral along a collection of small counterclockwise circles centered at $\zeta\in \mathbb{Z}/\{0\}$. By deforming the contour, it can be expressed as the residue at $\zeta=0$ and $\pm \alpha_0\equiv \pm \sqrt{-{24 \over c}b_0}$.}
\label{fig: contour2}
\end{figure}
As in the \diffs\ case, we express the soft mode contribution as a contour integral
\begin{align}
	&\cF(\chi)=1- {6\over \pi c} \oint_{\mathcal{C}} d\zeta { e^{i\zeta \chi}  e^{-{ i \pi  \zeta\over 2} } \over e^{- 2\pi i \zeta }-1 } {f_{-\zeta}(-{\pi\over 2} )f_{\zeta}(-{\pi\over2} )\over \zeta^2 \left( \zeta^2 +{24 \over c} b_0 \right)}\ ,\label{eq: integral for diffs}
\end{align}
where $\mathcal{C}$ is a collection of small counterclockwise circles around $\zeta\in \mathbb{Z}/\{0\}$. See Figure~\ref{fig: contour integral3}. Then, we will deform the contour to pick up the rest of the poles. For this, we need further assumptions on $f_n(-{\pi \over 2})$ that the integrand in \eqref{eq: integral for diffs} does not blow up at infinity, at least, for $\chi\in(-\delta,\delta)$ with a small positive constant $\delta$. Moreover, we assume that $f_{\pm \zeta}(-{\pi\over2} )$ have neither pole in the $\zeta$ plane nor zero at $\zeta=\pm\sqrt{-{24 \over c}b_0}$. Then, the deformation of the contour gives (See Figure~\ref{fig: contour integral4})
\begin{align}
	\cF(\chi)\,=\,1+{12 i\over  c}\sum_{n\in \mathcal{P}} \res_{\zeta=n}  { e^{i\zeta \chi}  e^{-{ i \pi  \zeta\over 2} } \over e^{- 2\pi i \zeta }-1 } {f_{-\zeta}(-{\pi\over 2} )f_{\zeta}(-{\pi\over2} )\over \zeta^2 \left( \zeta^2 +{24 \over c} b_0 \right)}\ ,
\end{align}
where we defined $\mathcal{P}\equiv \{0, \pm  \sqrt{-{24 \over c}b_0} \} $. Note that the pole at $\zeta=0$ would be simple pole because the time translational invariance of bi-locals imposes $f_0(\sigma)=0$. Even if $f_0(\sigma)\ne 0$, the pole at $\zeta=0$ does not change the Lyapunov exponent since it gives at most polynomial growth in time at $\mathcal{O}(c^{-1})$ order. Therefore, we have
\begin{align}
	\cF(\chi)\,=\,&1- {6 \over c} \cos \alpha_0(\chi +{\pi\over 2} )   {  f_{-\alpha_0}(-{\pi\over 2} )f_{\alpha_0}(-{\pi\over2} )\over \alpha_0^3 \sin \pi \alpha_0 }+ \cdots\ , \label{eq: four point function result s1}
\end{align}
where we define $\alpha_0\equiv  \sqrt{-{24 \over c}b_0}$ for simplicity. Here, the ellipsis represents the contribution of the pole at $\zeta=0$ which does not grow exponentially after analytic continuation. By the analytic continuation \eqref{eq: analytic continuation} of \eqref{eq: four point function result s1} from Euclidean time to Lorentzian one (See Figure~\ref{fig: oto}), the OTOC for \diffs\ is found to be
\begin{align}
	\cF(t)\,=\,1- {3 \over c} e^{{2\pi \alpha_0 \over \beta}t} e^{\pi i\alpha_0 \over 2}  {  f_{-\alpha_0}(-{\pi\over 2} )f_{\alpha_0}(-{\pi\over2} )\over \alpha_0^3 \sin \pi \alpha_0 }+ \cdots\ .
\end{align}
And, the Lyapunov exponent can easily be read off
\begin{equation}
	\lambda_L\,=\,{2\pi \over \beta}\alpha_0\,=\, \sqrt{-{24 \over c}b_0} \ .\label{eq: lyapunov exponent u1}
\end{equation}
For $b_0<0$, the bound on $b_0$ in~\eqref{eq: stability bound} for stability of the coadjoint orbit~\cite{Witten:1987ty} in large $c$ leads to
\begin{equation}
	\lambda_L\,=\,{2\pi \over \beta} \sqrt{{24 \over c}|b_0| }< {2\pi \over \beta}\ ,
\end{equation}
which corresponds to the bound on chaos proven in~\cite{Maldacena:2015waa}. For $b_0\geqq 0$, it is easy to see that the Lyapunov exponent is zero \ie $\lambda_L=0$.

\section{Conclusion}
\label{sec: conclusion}

In this work, we have evaluated the Schwarzian soft mode contribution to the OTOC for the case of \diffv\ and \diffs. We have showed that the stability of the coadjoint orbit found in~\cite{Witten:1987ty} leads to the bound of chaos proven in~\cite{Maldacena:2015waa}. While this bound is saturated by \diffv, the coadjoint orbit \diffs\ where the $\slt$ is broken to $U(1)$ is not maximally chaotic. \ie $\lambda_L< 2\pi/\beta$.

The decrease of the Lyapunov exponent has been observed in the generalized SYK-like models exhibiting the transition from chaotic phase to non-chaotic phase (or, from non-Fermi liquid phase to Fermi liquid phase)~\cite{Banerjee:2016ncu,Garcia-Garcia:2017bkg,Nosaka:2018iat,Ferrari:2019ogc}. The underlying mechanism for the decreased Lyapunov exponent in those models is the broken $SL(2,\mathbb{R})$ symmetry to $U(1)$~\cite{Ferrari:2019ogc}.\footnote{I would like to thank Frank Ferrari for pointing out this and for thorough discussions.} Furthermore, \cite{Anninos:2018svg,Mertens:2019tcm} studied the 2D dilaton gravity with defects where the $SL(2,\mathbb{R})$ isometry of AdS$_2$ background is broken to $U(1)$ due to the defect. This isometry of the background is responsible for the redundant description of the boundary modes, and it should be gauged. Therefore, the effective action for this case will be related to {\diffs} rather than \diffv. And one can also expect that the Lyapunov exponent does not saturate the bound (if the Schwarzian soft mode contribution dominates OTOC). In addition, this broken $SL(2,\mathbb{R})$ symmetry would essentially be responsible for the phase transition of the coupled SYK-like models dual to the traversable wormhole~\cite{Maldacena:2018lmt,Garcia-Garcia:2019poj,Kim:2019upg} where the phase transition is triggered by a similar (non-local) quadratic interactions.

In general, if the $SL(2,\mathbb{R})$ of a saddle point is broken to $U(1)$, all possible terms which vanish under $U(1)$ mode can appear in the effective action. For example, the simplest effective action can be written as
\begin{equation}
	S_{\text{\tiny eff}} \,=\, - \kappa_1 c \sch\left[ \tan  \left({ \phi(\tau) \over 2} \right),\tau\right]+\kappa_2 [\partial_\tau \phi(\tau)]^2\ ,
\end{equation}
where $\kappa_1$ and $\kappa_2$ are constants to be determined. In large~$c$, the stability of the semi-classical analysis, or equivalently, the chaos bound requires
\begin{equation}
	\kappa_2\,\geqq\, 0\ .
\end{equation}

From the Lyapunov exponent in \eqref{eq: lyapunov exponent u1} together with stability bound in \eqref{eq: stability bound}, one can also conclude that the instability of the coadjoint orbit action could lead to the violation of the bound on chaos. \ie $\lambda_L >2\pi /\beta$. An analogous phenomenon has been observed in the higher spin gravity~\cite{Perlmutter:2016pkf,Narayan:2019ove} and the Fishnet model~\cite{deMelloKoch:2019ywq}. The $SL(N)$ Chern-Simons higher spin gravity has finite numbers of higher spin fields (spin $s=2,3,\cdots, N$), and it is non-unitary. The Lyapunov exponent is found to be ${2\pi \over \beta}(N-1)$ which violates the bound on chaos for $N>2$. Moreover, the mass deformation of the Fishnet model destroys the integrability, but the Feynman diagrams are still simple. In this case, \cite{deMelloKoch:2019ywq} showed that the Lyapunov exponent can exceed $2\pi/\beta$ in the large 'tHooft limit, which the non-unitarity of the Fishnet model is also responsible for.

It is interesting to explore the role of the stability in the context of ``pole skipping'' phenomenon~\cite{Grozdanov:2017ajz,Haehl:2018izb,Blake:2018leo,Grozdanov:2018kkt,Grozdanov:2019uhi}. The pole-skipping, which is universal in two-dimensional CFT, can determine the Lyapunov exponent~\cite{Grozdanov:2017ajz,Haehl:2018izb,Blake:2018leo,Grozdanov:2018kkt,Grozdanov:2019uhi}. Therefore, one might be able to find a connection between the (in)stability of hydrodynamics and the bound on chaos via the pole-skipping. Furthermore, this might suggest a universal framework to understand how the (in)stability of hydrodynamical effective action leads to the bound on chaos in higher dimensional CFT.

\acknowledgments

I would like to thank Keun-Young Kim, Viktor Jahnke, Cheng Peng, Frank Ferrari, Sudip Ghosh and especially Robert de Mello Koch for extensive discussion. I thank the Kavli Institute for Theoretical Physics for generous support during the initial stages of this work, within the workshop ``Chaos and Order: from strongly correlated systems to black holes 2018''. This research was supported in part by the National Science Foundation under Grant No. NSF PHY-1748958. JY was supported by KIAS individual Grant PG070102 at Korea Institute for Advanced Study and the National Research Foundation of Korea (NRF) grant funded by the Korea government (MSIT) (No. 2019R1F1A1045971). JY is supported by an appointment to the JRG Program at the APCTP through the Science and Technology Promotion Fund and Lottery Fund of the Korean Government. This is also supported by the Korean Local Governments - Gyeongsangbuk-do Province and Pohang City. JY thank the Erwin Schrodinger International Institute (ESI) during the course of this work, within the program ``Higher Spins and Holography 2019''. JY thank the Okinawa Institute of Science and Technology~(OIST), the Gwangju Institute of Science and Technology~(GIST) and the South China Normal University~(SCNU) for the hospitality and generous support.

\appendix

\bibliographystyle{JHEP}
\bibliography{chaoscoadjoint04}

\end{document}